# Emergent Griffiths-phase-like behavior in the ball-milled nanocrystalline Dy$_4$RhAl and its implication

**Kartik K Iyer,$^1$ Kalobaran Maiti,$^{1,*}$ S. Rayaprol,$^2$ B A Chalke$^1$ and E.V. Sampathkumaran$^{3,*}$**

$^1$*Tata Institute of Fundamental Research, Homi Bhabha Road, Colaba, Mumbai 400005, India*

$^2$*UGC-DAE-Consortium for Scientific Research -Mumbai Centre, BARC Campus, Trombay, Mumbai 400085, India*

$^3$*Homi Bhabha Centre for Science Education, TIFR, V. N. Purav Marg, Mankhurd, Mumbai 400088, India.*

*Authors to whom all correspondences should be sent: sampathev@gmail.com; kbmaiti@tifr.res.in

**Abstract**

We report the results of dc susceptibility and heat capacity measurements on the (ball-milled) nanocrystalline rare-earth ($R$) ternary compound, crystallizing in Gd$_4$RhIn-type, cubic Dy$_4$RhAl compound (space group $F\bar{4}3m$, No. 216, cF96). The bulk form of this compound has been known to undergo antiferromagnetic ordering at ($T_N$=) 18 K with concomitant cluster spin-glass anomalies. The present studies on the nano-form obtained by ball-milling reveal that this antiferromagnetic ordering is suppressed with the reduction of particle size with no feature attributable to a well-defined long-range magnetic ordering down to 1.8 K, but showing an inhomogeneous magnetism below 10 K. The point being stressed is that the results show the dominance of a feature around 30 K in the magnetic susceptibility data - well above $T_N$ of the bulk form - mimicking Griffiths phase. We infer that surface magnetism dominates before long-range magnetic ordering occurs in this material.

{

One of the current trends in the field of magnetism is to look for various manifestations of magnetic frustration due to the competition between antiferromagnetic and ferromagnetic interactions in a material, attributable to geometrical arrangement of the magnetic ions in the lattice. It is now realized that a consequence of such a competition is the 'Topological Hall Effect,' the signatures of which were reported on $Gd_2PdSi_3$ [1], a decade before this terminology was applied to magnetic systems [2], and magnetic skyrmions [3, 4]. Considering the application potential of such systems, there are constant efforts to identify such exotic magnetic materials and to find ways and means of tuning the same. A very recent theoretical effort by Hayami [5] emphasized the need to focus on multi (magnetic) substructure systems to understand such a magnetic frustration. However, it should be noted that such multi-substructure systems among rare-earth (R) intermetallics are rather scarce, some examples being $Gd_5Si_2Ge_2$ [6] and $R_7Rh_3$ [7, 8]. In this respect, the compounds of the type $R_4TX$ ($T$ = Transition metal and $X$ = p block elements), which are characterized by three different crystallographic sites for the rare- earths [9-13] are of great interest, as the bulk specimens of the heavy rare-earth members $R_4RhAl$ ($R$ = Gd, Tb, Ho and Er) showed [14-18] exotic magnetic and transport properties. In the case of $R_4PtAl$ family, ($R$ = Gd, Tb, Dy, Ho, Er), all the members are known to order antiferromagnetically, with the Er and Ho members exhibiting a larger value of isothermal entropy change ($\Delta S$), at the onset of magnetic ordering within this family [19]. It may be mentioned that microscopic studies are not available at present to understand whether the observed anomalies in the bulk measurements are the result of the formation of magnetic skyrmions or any other modern magnetic concept like altermagnetism [20] (given that many of these compounds including the title compound are characterized by antiferromagnetic as well as ferromagnetic features and that the local symmetries for the three magnetic sites of R are different).

In this Letter, we focus on the compound $Dy_4RhAl$ [21], which has been reported to exhibit antiferromagnetic ordering at ($T_N$=) 18 K setting it concomitantly with spin glass features, apart from

{

other subtle magnetic anomalies as a function of temperature (*T*) and magnetic field (*H*). The study on fine particles was motivated by the fact that some of the rare-earth intermetallics showed interesting changes in properties when going from bulk to nanoform including conversion of Pauli paramagnetism to magnetic ordering, e.g., $YCo_2$ [22], and a well-known heavy-fermion $CeRu_2Si_2$ [23]. This work is further motivated by the importance of understanding magnetism of muti-substructures systems in different forms. The results reveal suppression of antiferromagnetism in the nano form, leading to a Griffiths-phase-like state – a magnetic state proposed for a situation in which ferromagnetic clusters are distributed in a paramagnetic state [24-28] – in the *H* dependent $\chi(T)$ data. We argue that there is a qualitative change in the magnetic properties when traversing from the bulk to the surface of complex magnetic materials, particularly the ones with multiple magnetic sites.

A polycrystalline sample of $Dy_4RhAl$ was prepared by melting together stoichiometric amounts of constituent elements in an arc furnace in an argon atmosphere. The phase and stoichiometry of the ingots were ascertained using x-ray diffraction (XRD) and Energy Dispersive Analysis of x-rays (EDAX). The ingot was ground using a high energy ball mill (Pulverisette 7, M/s. Fritsch GMBH, Germany) for 150 minutes at a speed of 500 RPM in a zirconia bowl with zirconia balls of 5 mm diameter. A small amount of milled sample was mixed with a drop of diluted GE-Varnish and allowed to dry to get a single pellet. This pellet was used for measurements. A transmission electron microscope (TEM, Technai-200kV) was used to characterize the phase and the particle size. Dc $\chi$, isothermal magnetization (*M*) and heat- capacity (*C*) measurements were performed as a function of *T* and *H*, as described in Ref. 21.

Fig. 1a shows the XRD data for the powders of molten ingot and the ball-milled materials. The main panel of the figure shows the profile fitted using Rietveld refinement method along with the observed powder XRD data for the bulk sample. The fit clearly shows the formation of the sample in $Gd_4RhIn$-type $Dy_4RhAl$ cubic structure (space group $F\bar{4}3m$, No. 216, cF96), with no additional peaks.

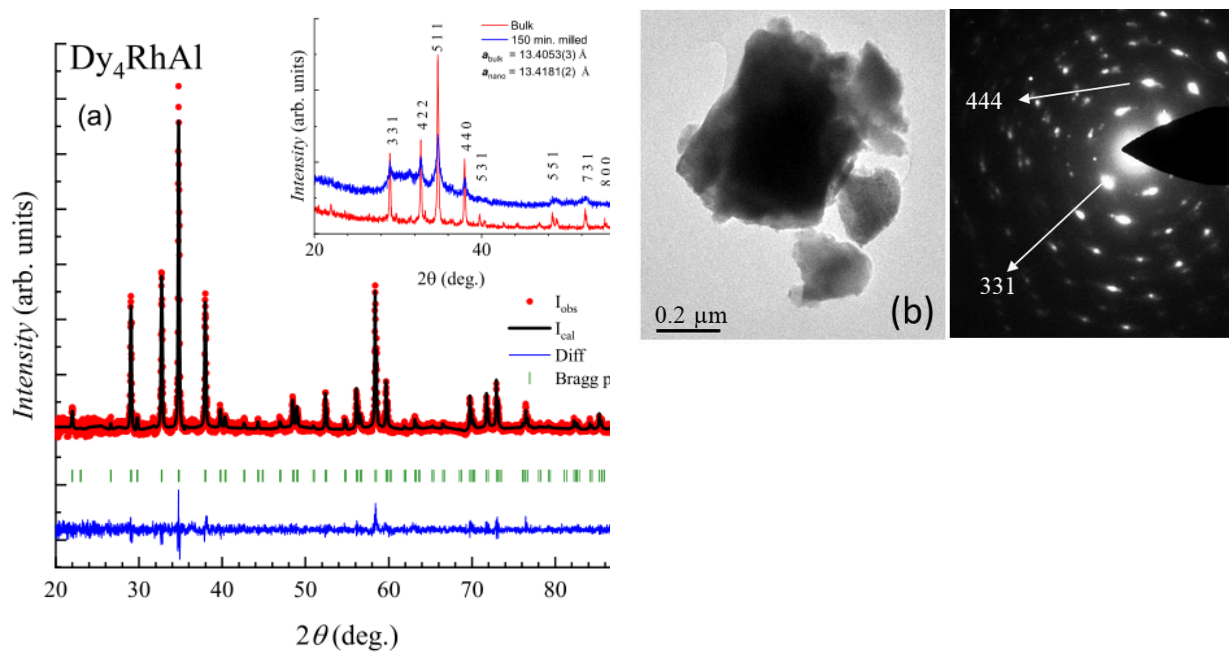

{

**Fig 1**. (a) Powder x-ray diffraction pattern for the bulk sample fitted to a Rietveld refined structural model is shown in the main panel. The plot shows the observed data (Iobs), calculated profile ($I_{cal}$) and the difference profile ($I_{obs} - I_{cal}$) along with the Bragg peak positions shown as vertical tick marks. (The inset shows powder X-ray diffraction patterns for bulk and nanoform specimen (obtained by ball milling for 150 minutes) of $Dy_4RhAl$ at room temperature. (b) TEM images to reveal particle sizes and (c) electron diffraction pattern obtained on a single nano particle to establish phase formation.

The peaks in the XRD pattern for the milled sample show broadening. We calculated the particle size for this specimen from the XRD measurements by Debye-Scherrer formula and the average particle size was found to be in the range of 20 - 50 nm. Table 1 summarizes the values of refined cell parameters. The transmission electron microscopic pictures, shown in Fig. 1(b), show the presence of nanoparticles of higher sizes as well (a few up to 500 nm), suggesting an inhomogeneity in particle sizes, which is inevitable in the ball-milling procedure. The electron diffraction pattern shown in Fig.1(c) was obtained on a particle of about 200 nm and confirms that there is no change in the structure after milling (as inferred from the indexing of some of the diffraction lines).

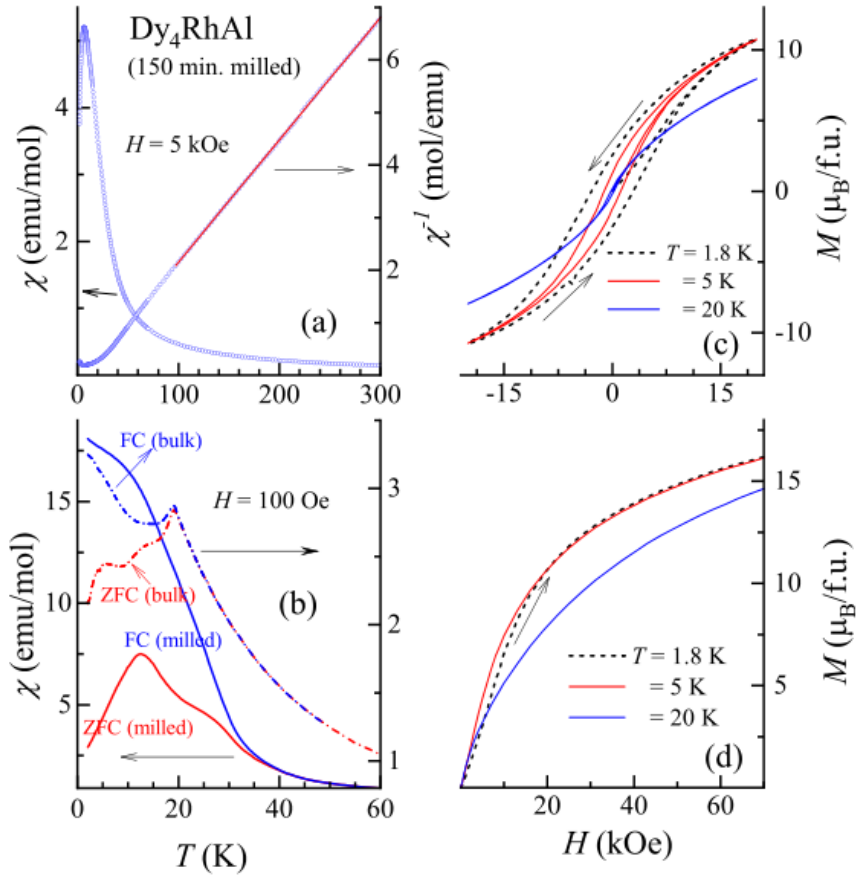

**Fig. 2.** Temperature dependent dc susceptibility measured in (a) 5 kOe (zero field cooled condition) and (b) 100 Oe (for zero field cooled and field cooled conditions) for milled specimens. The curves for the bulk form are reproduced from Ref. 20. (c) Magnetic hysteresis loops at 1.8, 5 and 20 K. (d) Isothermal magnetization in units of Bohr magneton per formula unit (f. u.) at selected temperatures measured up to 70 kOe.

{

We now focus on the effects of the reduction in particle size on the magnetism in this compound. In Fig. 2(a), we show the $\chi$ and inverse $\chi$ data measured in a magnetic field of 5 kOe down to 1.8 K for the nanoform specimens; inverse $\chi$ remains linear down to 50 K and obeys Curie-Weiss behaviour. The paramagnetic Curie temperature ($\theta_P$) obtained for the linear region is close to 7.6 K, with the positive sign indicating effective ferromagnetic correlations in contrast to the negative sign of $\theta_p$ (= -16 K) for the bulk form typical of antiferromagnetic interactions [21]. Table 1 gives a comparison of the values of $\theta_p$ and $\mu_{eff}$ for bulk and nano samples. A lower value of effective moment for milled specimen is attributed to difficulties in estimating mass of the sample in the presence of GE varnish component.

**Table 1:** Values of unit cell parameters obtained from the refinement of powder diffraction data is summarized here. The table also lists the values of paramagnetic Curie temperature and effective magnetic moment obtained from the magnetic susceptibility data for bulk and nanosized samples of $Dy_4RhAl$.

| Sample | Cell parameter ($a$, Å) | Paramagnetic Curie temperature ($\theta_p$) | Effective magnetic moment ($\mu_B$) per formula unit |
|---|---|---|---|
| Bulk | 13.4053(3) | -16 K | 21.9 |
| Nano | 13.4181(2) | 7.6 K | 18.5 |

As the temperature is lowered, there is no feature around 18 K attributable to the magnetic ordering seen in bulk specimen, but we find a distinct peak appearing at 8 K in this 5kOe-curve. It is not clear whether it is the same additional feature seen in the bulk specimen around 10 K, when measured with 5 kOe. As seen in Fig 2(b), for a magnetic field of 100 Oe, $\chi$ shows a prominent bifurcation in the ZFC-FC curves in the vicinity of 30 K for the milled specimen, which is absent in similar measurements on the bulk form, as shown in the same figure. However, note that even in the bulk form [21], there is a weak bifurcation of the two curves, obtained by measuring with 100 Oe and 5 kOe, well before magnetic ordering sets in, as though there is a weak magnetic anomaly above $T_N$. In addition to a shoulder around 30 K in this ZFC curve, we see a broad peak appearing around 12K which is similar to the shoulder around 12 K for the bulk specimen in the data measured with 100 Oe, except that this shoulder has become relatively more prominent and broadened in the nanoform specimen. Finally, the FC curve, rather than remaining flat at the temperature of bifurcation from the ZFC curve, keeps increasing with decreasing temperature, typical of cluster spin- glasses.

In order to throw more light on the changes observed in the magnetic ordering behaviour, the heat capacity results are shown in Fig. 3(a) and 3(b), along with the zero-field data [reproduced from Ref. 21] for the bulk specimen. We find that $C$ varies monotonically down to 1.8 K, without any evidence for a $\lambda$-anomaly, seen prominently for the bulk specimen around 18 K, supporting the absence of long-range antiferromagnetic ordering in nanoform specimens.

{

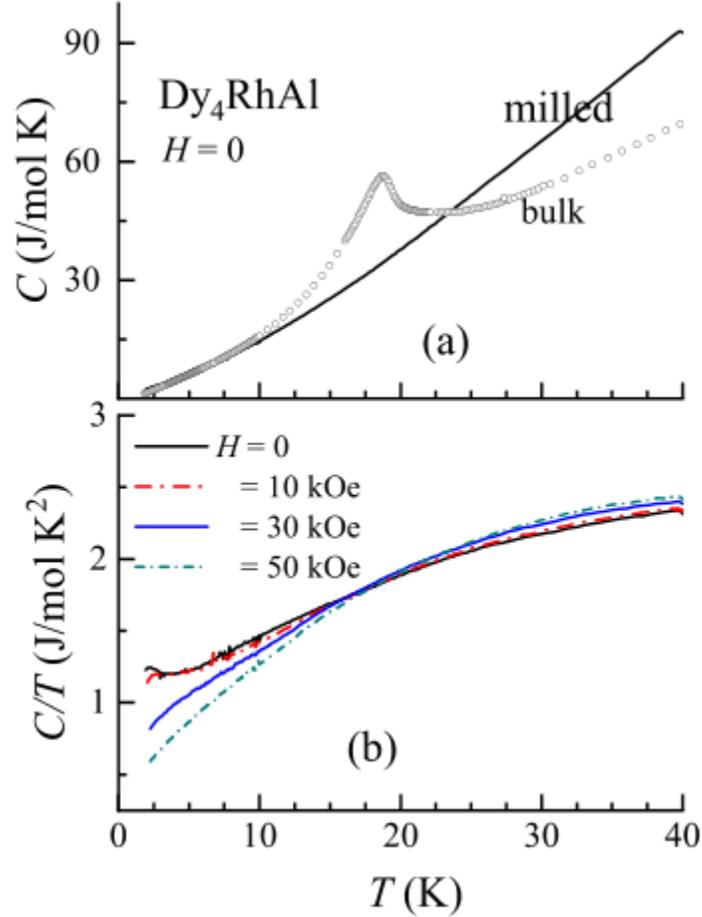

**Fig. 3.** Plots of (a) heat capacity (C), and (b) heat capacity divided by temperature as a function of temperature (in the presence of 10, 30 and 50 kOe) in the vicinity of magnetic ordering. The heat-capacity data for the bulk (Ref. 20) is also shown in (a).

We find that the $C(T)$ data for 30 kOe and 50 kOe almost overlap with the $H = 0$ data down to 1.8 K (and hence not shown in Fig. 3(a) for the sake of clarity). However, a careful look at the $C(T)/T$ plots (Fig. 3(b)) shows that, there is a subtle variation of the heat capacity values for a field of 30 kOe and 50 kOe, which points to the persistence of some sort of magnetism. Incidentally, we have also derived isothermal entropy change [$\Delta S=S(H)- S(0)$] from the heat-capacity data by integrating $C/T$ over temperature and the sign of the $\Delta S$ remains negative with a broad peak, similar to that noted for the bulk form [21] (and hence not shown here); this negative sign is a signature [29] for a tendency for the field-induced ferromagnetic alignment. Combined with the fact that the bifurcation in the $\chi(T)$ curves begins in the range 30 to 35 K for $H= 100$ Oe, the $C(T)$ data provides evidence for the inference that the glassy nature gets enhanced, and the long-range antiferromagnetic order gets suppressed when the sample is driven to the nanoform in this compound.

We present the $M(H)$ data for the nanoform specimen in Figs. 2(c) and 2(d). We find that the $M(H)$ curves are distinctly hysteretic at 1.8 and 5 K [in the range (-20) to 20 kOe] supporting the existence of a spin-glass-like or ferromagnetic component. However, the hysteresis is negligible at 20 K. The (virgin) curves extended to higher fields are shown in Fig. 2(d). There

{

is no evidence for saturation even for a field as high as 70 kOe, which is typical of spin-glass and antiferromagnetism. Arrott plots (Fig. 4) offer an additional insight into the low temperature magnetism. These plots - $H/M$ versus $M^2$ - are shown at different temperatures for both the bulk form as well as milled form. The slope of the virgin curve (that is, for the upward cycle of field variation) remains positive for temperatures well above 10 K, whereas a negative slope appears in the virgin curves for $T = 2$, 5 and 10 K. This is true for the bulk form also. This signals [30-31] that the zero-field magnetic state at such low temperatures, presumably containing some antiferromagnetic component (as inferred in Ref. 21), undergoes a disorder-broadened first-order field-induced transition to a ferromagnetic-like component. The fact that, in the return cycle, the slope remains positive for the milled form after a high magnetic field cycling is in support of a first-order field-induced transition for the virgin state only.

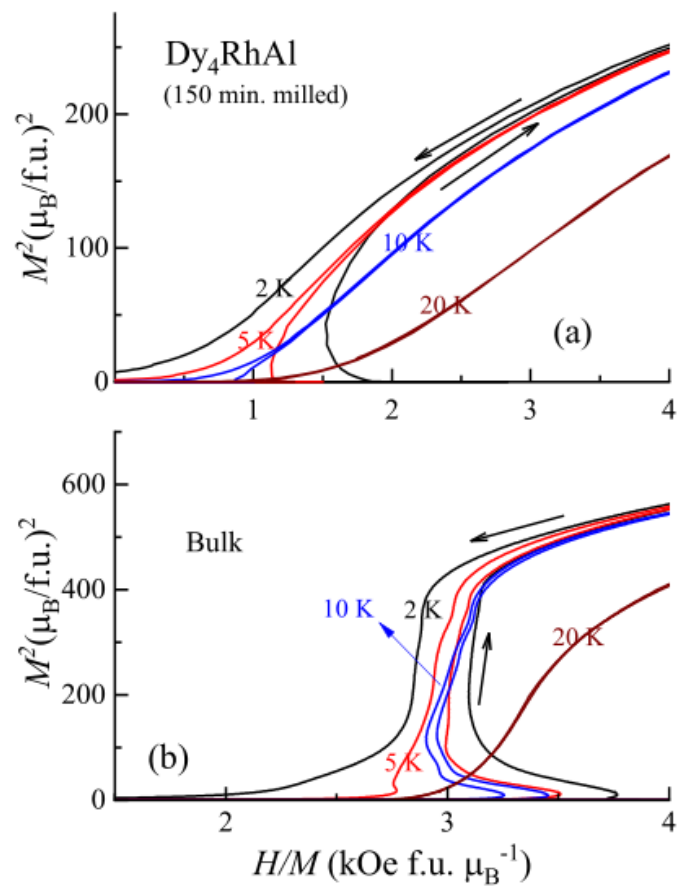

**Fig. 4.** Arrott plots for the bulk and milled forms of $Dy_4RhAl$ at the temperatures specified. Arrows are drawn to show the direction in which the $H/M$ increases

This behavior is different from that of the bulk, as the negative curvature is retained in the reverse cycle of field variation as well. Thus, there are also subtle changes when reducing the particle size. Finally, a careful look at the $M(H)$ curves at 2 K, for both bulk and milled forms, shows a feeble upward curvature at low-fields (<4 kOe) typical of an antiferromagnetic component in zero-field, which is smeared in the 5 and 10 K curves (Fig. 2(c)). Thus, there are features attributable to spin-glass-like, ferromagnetic-like as well as antiferromagnetic behavior

{

at very low temperatures (< ~10 K). It is not clear at present whether this apparent inhomogeneous magnetic state is due to a distribution in particle size, a surface phenomenon or it implies a more complex magnetism for small particles.

To get an insight into field-dependent behavior of magnetism of the specimen above 10 K – which is the central point of this Letter - we have performed the χ(T) measurements for different applied magnetic fields, the results of which are shown as $\chi^{-1}(T)$ plots in Fig. 5(a). The curves above 50 K tend to overlap. We find that the feature below about 10 K (a broad upturn) shifts towards a low temperature marginally with the increase in the magnetic field initially, except for a magnetic field of 5 kOe, where we find a sharp drop of temperature from 8.5 K for $H = 3$ kOe to 5 K for $H = 5$ kOe. This trend continues till this characteristic temperature reaches 3K for a magnetic field of 20 kOe, beyond which we do not find any such worthwhile feature in the $\chi^{-1}(T)$ as the magnetic field is increased.

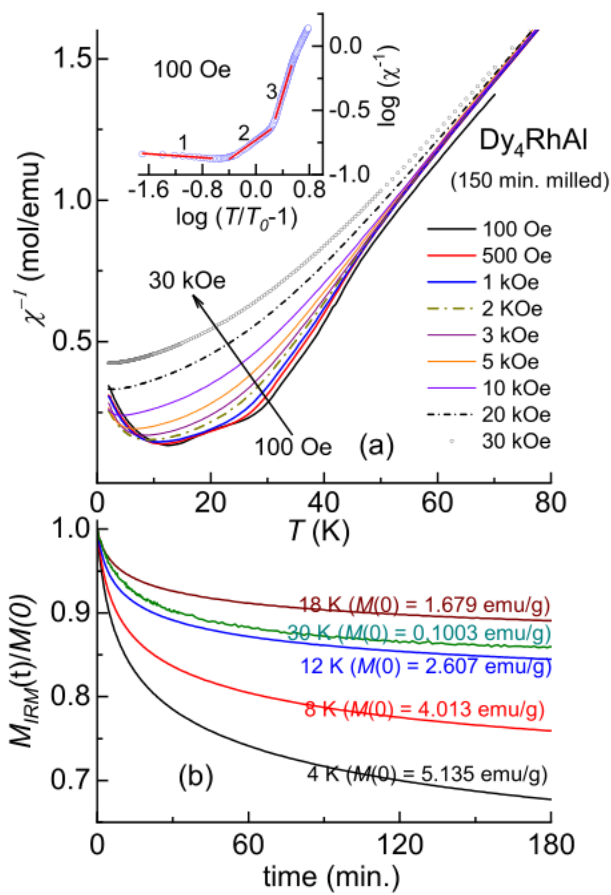

**Fig. 5**. (a) Inverse magnetic susceptibility as a function of temperature in the presence of dc magnetic fields. Inset shows data for 100 Oe, plotted as log ($\chi^{-1}$) verses log($T/T_0$-1). The three red lines are the results of a linear fit in regions marked, as 1, 2, and 3, (b) Isothermal remnant magnetization curves obtained at several temperatures, and the lines through the data points obtained by a fit as described in the text; the values are normalized to the respective initial values ($M_0$) immediately after switching of the field.

{

This may be correlated to the anomalies seen at such low temperatures in the $\chi(T)$ data discussed above (Fig. 2(b)). *The main point we emphasize is that there is a sudden change of slope around 30 K in the low-field curves, which is gradually smeared with increasing H.* Qualitatively speaking, such a dependence of the $\chi^{-1}(T)$ with applied magnetic fields is the fingerprint of Griffiths Phase. Such a magnetic phase was conceptualized long ago [24-28] to explain a situation where traces of ferromagnetic clusters are distributed randomly in a paramagnetic matrix in such a manner that its contribution to $\chi$ is overshadowed by paramagnetic contribution. Thus, it is generally not easy to detect by magnetic studies. However, this magnetic behaviour, mimicking Griffiths phase has been recently claimed to occur in dense magnetic systems also, more frequently among oxides [see, for instance, Refs. 32-34]. Many Ce based intermetallics, characterized by extended 4$f$ orbital, near the quantum critical point exhibiting non-Fermi liquid behaviour have also been shown to exhibit the characteristics of this phenomenon [28, 35-37]. But it is less commonly demonstrated in stable valent rare-earth intermetallics with well-localized 4$f$ orbital [38]. It is in this context that, for a heavy rare-earth compound with strictly localized 4$f$ orbital showing antiferromagnetism in the crystalline form, the observation of Griffiths-like behavior in the ball-milled nano form is interesting. To further ascertain that the features observed in Fig. 5(a) signal Griffiths phase, we have looked for the $T$-region where the theoretical prediction [26-28] is obeyed. That is, $\chi^{-1}(T)$ should be proportional to $(T-T_0)^{1-\lambda}$ where $\lambda$ should be in the range 0 to 1. The plot of $\chi^{-1}$ versus $(T/T_0-1)$ is shown in Fig. 5(a) (inset) for the data measured in 100 Oe. We can identify three linear regions in this plot, as shown in this figure. The values of $\lambda$ derived from the slopes are: ~ 1 (~10.2 to ~12 K), ~ 0.6 (~13 to ~ 27 K), ~ - 0.7 (~ 30 K to ~ 45 K) On the basis of this, one can conclude that approximately in the range 10 K < $T$ < 30 K only, the value of $\lambda$ turns out to be positive and less than 1. The value of $T_0$ is 10 K. In the linear region above 30 K, the sign of this exponent is unrealistically negative. Thus, the results show the signature of Griffiths phase in this compound below 30 K when the particle size is reduced to nanoform, before a different kind of inhomogeneous magnetic state evolves below 10 K. We believe that non-overlapping curves for different $H$ in Fig. 5 at higher temperatures is due to conventional short-range magnetic correlations (given that $\lambda$ value does not obey Griffiths formula), which are often encountered in many bulk magnetic materials before long-range magnetic order sets in.

We have also measured isothermal remnant magnetization ($M_{IRM}$) as a function of time ($t$) at selected temperatures below 30 K. This measurement was performed first by cooling the sample in the magnetometer from 100 K to a desired temperature, thereafter, leaving in a field of 5 kOe for about 5 mins; the field was then switched off and subsequently $M_{IRM}$ behavior was tracked as a function of $t$. The curves thus obtained are shown in Fig. 5(b) and it is clear that there is a slow decay of $M_{IRM}$, suggesting that the Griffiths-phase-like regime (see the curves for 12, 18 and 30 K) exhibits spin-glass dynamics [39, 40]. This slow decay persists down to 1.8 K, attributable to glassiness. The decay curves are also found to obey a stretched exponential form, $M_{IRM}(t) = M_{IRM}(0) + A \exp(-t/\tau)^{1-n}$, where A is a constant, and the time constant $\tau$ and the exponent n are related to the relaxation rate of the clusters.

To summarize, the present fine particles (<500 nm) studies of $Dy_4RhAl$ reveal that the antiferromagnetic ordering dominating in the bulk specimen, disappears in the nanocrystalline form, while a new magnetic state emerges in the range 10-30 K. Considering that this feature is feebly present for the bulk form as well below ~35 K, visible in some form when measured with

{

different low fields, we attribute this magnetic state to the surface layers (naturally characterized by disorder due to incomplete coordination or strain) since the total area of the surface gets enhanced when the particle size is reduced. The magnetism of such a state is characterized by a magnetic susceptibility behavior mimicking Griffith's phase around 30 K. Demonstration of such a particle-size induced transformation to Griffith's phase-like features in an intermetallic compound due to localized 4f electrons is generally quite rare. It is of interest to explore how common this is in other exotic magnetic systems.

Authors acknowledge financial support from the Department of Atomic Energy (DAE), Govt. of India (Project Identification no. RTI4003, DAE OM no. 1303/2/2019/R&D-II/DAE/2079 dated 11.02.2020).thanks Department of Atomic Energy, Government of India, for awarding Raja Ramanna Fellowship. We thank Jayesh Parmar for his help during characterization by TEM.

{

{